\documentclass{article}

\PassOptionsToPackage{square,sort,comma,numbers}{natbib}
\usepackage[preprint]{nips_2018}

\usepackage[utf8]{inputenc} % allow utf-8 input
\usepackage[T1]{fontenc}    % use 8-bit T1 fonts
\usepackage{hyperref}       % hyperlinks
\usepackage{url}            % simple URL typesetting
\usepackage{booktabs}       % professional-quality tables
\usepackage{amsfonts}       % blackboard math symbols
\usepackage{nicefrac}       % compact symbols for 1/2, etc.
\usepackage{microtype}      % microtypography
\usepackage{adjustbox}
\usepackage{array}
\usepackage{color}
\usepackage{xcolor}

\setcitestyle{square}

\newcolumntype{R}[2]{%
    >{\adjustbox{angle=#1,lap=\width-(#2)}\bgroup}%
    l%
    <{\egroup}%
}

\title{A Survey of Mobile Computing \\for the Visually Impaired}
\author{
  Martin Weiss$^{\ddagger}_{\ast}$  \And
  Margaux Luck$^{\ddagger}_{\ast}$   \And 
  Roger Girgis$^{\ddagger}_{\triangleright}$ \And
  Chris Pal$^{\ddagger}_{\diamond}$   \And
  Joseph Paul Cohen$^{\ddagger}_{\ast}$  \And   \vspace{-10pt}\\
  $^{\ddagger}$ Montreal Institute for Learning Algorithms, $_\ast$Université de Montréal \\
  $_{\diamond}$Polytechnique Montréal, $_{\triangleright}$McGill University\\
}

\begin{document}

\maketitle

\begin{abstract}
    %We present BlindTool, a mobile application for the visually impaired built with an Apache MXNet object classifier and a novel human-computer interaction pattern for communicating with the user. This prototype application was released to the public in 2016 and has received media significant attention, more than 10,000 downloads, and hundreds of pieces of feedback. We provide analysis of the feedback with a focus on the successes and failures of BlindTool and the limitations of on-device machine learning.
    
    The number of visually impaired or blind (VIB) people in the world is estimated at several hundred million\cite{bourne2017magnitude}. Based on a series of interviews with the VIB and developers of assistive technology, this paper provides a survey of machine-learning based mobile applications and identifies the most relevant applications. We discuss the functionality of these apps, how they align with the needs and requirements of the VIB users, and how they can be improved with techniques such as federated learning and model compression. As a result of this study we identify promising future directions of research in mobile perception, micro-navigation, and content-summarization.
    %listing the areas of research that could yield the greatest impact on the lives of the visually impaired.

\end{abstract}

\section{Introduction}

    The World Health Organization estimates that there are 253 million visually impaired or blind (VIB) people in the world~\cite{bourne2017magnitude}. These people typically use a white cane to help them navigate the world, enabling them to detect uneven floors and static obstacles. Guide dogs can help VIB people detect and avoid dynamic obstacles such as cars, bicycles, and pedestrians. Nevertheless, common tasks such as sorting mail and medications, reading, separating laundry, selecting items on a menu, or navigating unknown environments still present a tremendous challenge.

    Recent advances in machine-learning and mobile technology are starting to pave the way towards a higher degree of autonomy for the visually impaired through the development of software that can identify objects, extract text, or describe scenes. Organizations like Aira.io and BeMyEyes or initiatives like VizWiz:Social~\cite{brady2013visual} provide mobile applications that connect VIB users to sighted volunteers or employees through asynchronous image captioning and live video calls to handle more complex tasks. 
     
    This paper presents the existing mobile application landscape for the visually impaired. We discuss currently available features, the successes and failures of those technologies, and the various machine-learning models they employ. 
   % By investigating apps that pair VIB users with sighted volunteers and professionals, we identify areas where automated solutions fail.
   We conclude by presenting our vision for the creation of better tools and where research should focus to help people with visual impairments.

    % A similar application deployed by researchers from Microsoft and the University of Rochester, VizWiz Social, also pairs VIB users and volunteers. It allows VIB users to take pictures of objects and speak a question about the contents of the picture, which the volunteer then answers~\cite{brady2013visual}. After collecting over 40,000 image question pairs, the researchers developed a set of 4 categories that the pairs fell into: Identification, Description, Reading, and Other. Identification questions were the most common, making up 41\% of the pairs. The researchers also grouped the question-image pairs by perceived urgency, with more than two thirds of the pairs placed in a high urgency category, requiring a reply within 10 minutes.

    % Answer here what kinds of things are important features for the visually impaired.
    
    %Other machine-learning-based solutions like SeeingAI, Envision AI, KNFB reader use technologies like micro-navigation, model compression, document recognition, ... to enable new levels of autonomy never before afforded to the VIB helping them in the orientation and mobility in their residences or at the required places with some assistive features of the system.

\section{Features of Assistive Tools}
\begin{figure}[h]
\centering
\includegraphics[width=\textwidth]{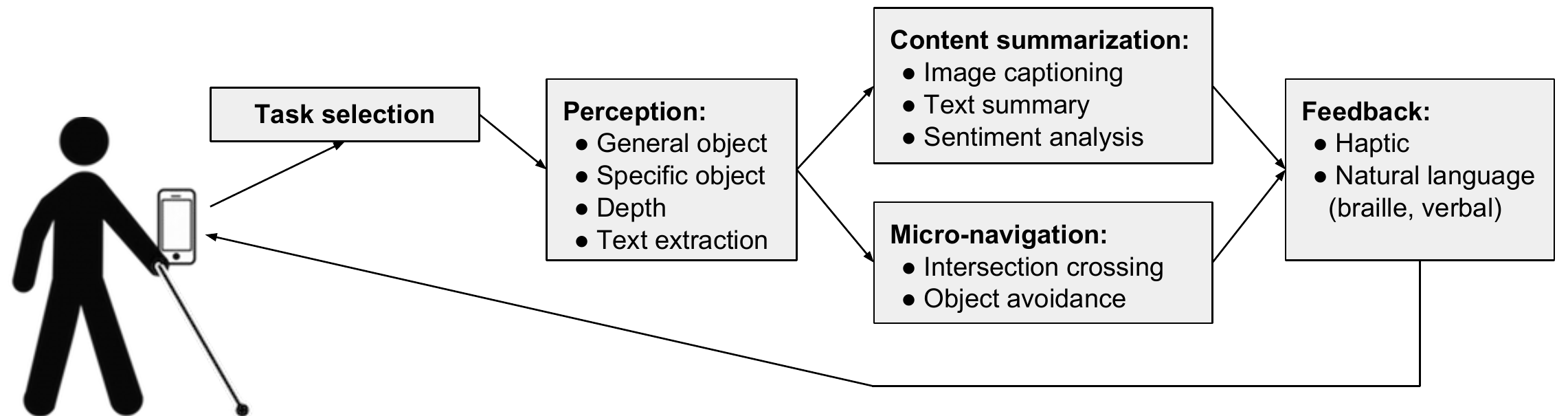}
\caption{The conceptional framework for most assistive tools for the VIB.}
\label{blind_tools}
\end{figure}

Visually impaired or blind people would greatly benefit from increased autonomy. The software thus aims to provide them with opportunities and abilities that rival those of the sighted. This requires a robust set of features that includes object detection, robust image captioning, text extraction, semantic understanding, and micro-navigation. Figure \ref{blind_tools} shows the conceptual framework for most assistive tools for the VIB. This should be used to guide the further development of tools and explore what is conceptually missing.

\subsection{Perception}
% Need a paragraph here

\textbf{General object detection} is necessary for daily activities like the localization and identification of furniture, computers, and food or drug containers. While a common feature, its actual effectiveness varies greatly across apps.  For example, BlindTool \cite{Cohen2015} (2015) has the ability to quickly detect objects using an on-device neural network, but is restricted to only a thousand classes. A problem with all current object detectors is generalization. For example, apps like NantMobile Money Reader, LookTel Money Reader, Eye Notes, and others deliver high performance identification on a range of currencies, but they fail to generalize to newly distributed versions of bills.   %Failing at content summarization: People generally do not wish to identify bills individually, they want to know how much money they received in change. 

%The VIB user can scan a room in order to detect objects. The application guides the VIB to a specific location to better detect certain objects.

% EnvisionAI uses a modified version of mobilenet for object segmentation and detection. Uses coreml. Trained on open images dataset from google, and retrain it "when they see new data".
% ThirdEye, TapTapSee, BlindTool, Minerva
% Barcode Scanning

%\subsubsection{Specific Objects (Friends and Possessions)}
\textbf{Specific object detection} refers to the task of identifying specific objects rather than general classes. These primarily include either objects familiar to an individual user such as their keys, car, wallet, or even friends, or those that are common knowledge such as famous landmarks or individuals. For example, Talking Goggles can recognize a wide variety of books, movies, and celebrities, whereas SeeingAI can fine-tune a device's object detection model on relevant user-generated images so that the user can recognize unique classes. These functionalities are a fundamental step towards higher-order tasks such as micro-navigation and scene description. 

% Need another paragraph here. Primarily wants to talk about EyeCYou's usecase where they discuss people's physical attributes. Talk about body language. Talk about SeeingAI recognizing your wallet. Discuss privacy too.

%Discuss SeeingAI and EnvisionAI as examples of applications that enable blind users to recognize their friends and personal possessions, and how they both use different methods to accomplish this. Discuss the privacy implications of training the network on a new person or object and how it's good that this is done on mobile. Question how useful this type of functionality is without greater context (What they want: Jimmy got a hit and is running to first base vs. what they get with EyeCYou's usecase where they give you a physical description of the person). It's an important piece of infrastructure, but doesn't really provide value alone.

% trained on device to recognize people and specific objects. Take 10 images, and doesn't work on iPhone 5S and earlier

%\subsubsection{Depth}
\textbf{Depth}: Thanks to stereo vision, the sighted can estimate the depth of their surroundings quickly and reliably. The app Seeing With Sound \citep{10.1109/10.121642} provides a proxy for this ability by translating an image into sound that the user can interpret. By moving the device around, the user can view the scene from different directions and determine the approximate depth. The challenge lies in acquiring a suitable depth map through for instance a stereo camera or extrapolating from a single camera's output, and to then communicate its complexity to the user. One potential approach is to simply present a dynamic 3D representation of the scene to the user but no such solution has been developed yet.

%\subsubsection{Text extraction}
\textbf{Text extraction} technology enables common tasks such as reading labels, price tags, documents, or pieces of mail and making sense of linear formats. The VIB user is prompted to position the text in front of the camera and then listen to audio feedback to center or scan the full text. Apps based on optical character recognition (OCR) such as KNFB Reader are rich in features. They allow the user to navigate by line, sentence, word, or individual characters in a wide variety of languages. Apps like Minerva, SeeingAI, and EnvisionAI also offer on-device OCR, although it is not their core feature. All currently available apps struggle to parse visually oriented layouts.

%OCR +. KNFB Reader is very fully-featured. Requires you to take a photo with the document centered, but has tools to help you get the document centered. Costs 100 dollars. Navigates by line, sentence, word, or character. Can read labels and price tags. Struggles with mail and arbitrary tables. Handles 15 languages fully, and another 17 for speech and recognition only.  Possibly mention the other apps listed below which also offer on-device OCR, although with poorer feature-sets. Also discuss the problem of understanding visually oriented layouts and the difference between what a model with a semantic understanding of the document could do (Q/A) and how you have to navigate an OCR'd document.

% Document Understanding (tables, question answering, SQuAD) 
% KNFB Reader, Minerva, SeeingAI

% EnvisionAI
% On device, only a subset of latin based languages. They are working on adding hebrew. Quantize the model. Most used service is text recognition which relies on open datasets.

\subsection{Content Summarization}

\textbf{Image captioning} solutions use natural language to describe a scene. Querying a semantic representation of the environment can yield contextually relevant information to the VIB user. Unfortunately, available applications like BlindSight which perform the image captioning task on-device suffer from low accuracy, whereas apps like SeeingAI and EnvisionAI which offload the task to a server suffer from slow response times and sub-optimal interaction patterns. The lack of semantic and contextual representations of information in neural networks presents a major obstacle for designers of assistive technologies that provide image captions, except for BeMyEyes and VizWiz which crowd-source captions from human volunteers or employees.

% you shouldn't have to take a picture for the network to say "There is a colorful bus coming up the street" when you are waiting for the bus. https://cs224d.stanford.edu/reports/mcelamri.pdf
%Also called scene description, this task return to the visually impaired user a natural language description of the scene before them.
%Scene Description task. Ideally in narrative form: https://arxiv.org/pdf/1611.07810.pdf

\textbf{Text summaries} require a semantic understanding of the contents and context of a document. Significant amounts of prior information are often needed. Several apps, like Summly, Trimit, and WrapItUp have advanced technology for the summarization of news articles, papers, and other documents. However, we have failed to find any app which contains both text summarization and OCR functionality, a combination necessary for the interpretation of physical documents. These technologies cannot yet reliably extract, summarize, or make sense of tabular data.

\textbf{Sentiment analysis} is a necessity for inter-personal interaction. While the tone and semantic content of spoken words carries significant amounts of information, VIB people must rely solely on these. Computer vision-assisted analysis can identify non-verbal cues such as facial expressions and posture, but gestures and significant eye movements remain challenging to identify. Apps like SeeingAI, EyeCYou and Minerva achieve this task off-device mainly by using APIs such as SkyBiometry and Microsoft Azure cognitive services.

% {\color{blue}RG: Some previous work explored blind people's interaction with social media, which has become filled with visual content. A lot of work on Twitter from Microsoft \cite{salisbury2017toward, macleod2017understanding, wu2017automatic}. Not really mobile, I guess, but some lessons learned are transferable.}

\subsection{Micro-Navigation}
We define the micro-navigation task as that of traversing the last hundred meters to a destination. Aira.io solves this problem by pairing sighted professionals with VIB users through smart-glasses and synchronous video chat.
%For example, if a user needs to get to a restaurant towards which the GPS cannot guide them due to measuring errors, they can rely on a sighted assistant who can survey the situation from the glasses' first person perspective.
For some VIB people, hiring professional assistants is prohibitively expensive, with plans ranging from \$.50-\$1.00 per minute. Additionally, smart-glasses often lack cosmetic acceptability, are expensive, and have limited battery life.  %, and are generally not designed for use by the visually impaired, although Aira.io has released a dedicated pair of smart-glasses for their userbase. 
 Automating this task will require robust localization, path-planning, intersection crossing and obstacle avoidance. % and human-computer interaction.
% TODO: Discuss path planning

%\textbf{Localization} and visual mapping are tasks that provide the user with an estimation of their precise location. Commonly used in augmented reality and autonomous vehicle applications, precise localization within the environment provides context to the assistive tool to improve micro-navigation accuracy. % Landmark extraction, data association [https://dl.acm.org/citation.cfm?id=378053], and state estimation supplement the GPS and IMU sensors in their smart-phone.
% I vote that localization is a perceptual task and should be in the perceptual tasks box, not the navigation box.

\textbf{Intersection Crossing} is considered the most difficult and risky aspect of independent navigation for the VIB people~\cite{shen2008mobile}. To safely accomplish this task the user must be aware of the spatial configuration of the intersection, the intersection's signalling pattern, the correct orientation to follow while crossing, and the time when it is safe to begin crossing. While previous research has tackled these problems individually, we are not aware of an available mobile application capable of assisting with all the tasks involved in this challenge.

\textbf{Obstacle Avoidance} is primarily handled by the traditional assistive tools, such as guide-dogs or canes. While previous work has proposed solutions to tackle obstacle avoidance using machine learning and vision~\cite{Poggi2016AWM}, their solution remains restricted to the lab setting as they require specialized hardware that is not commercially available. Options like the iOS BeAware app use a combination of beacons planted in the environment and bluetooth and WiFi. We are not aware of a viable commercial mobile application that tackles obstacle avoidance.

\section{Technical Details}
There are several obvious drawbacks to the apps we just presented: they can be expensive, lack cross-platform or multi-language support, have high latency, and are not always available. This is partly due to the fact that internet access is not always available raising the question of local versus remote processing (on- or off-device). Also, for some specific features, embedded models are needed to guarantee privacy.  To achieve the trade-off between privacy, accuracy, autonomy and accessibility those apps use light models with specialized architecture and also machine learning techniques like one-shot object detection, model compression, and federated learning.

\subsection{Specialized Architectures and One Shot Object Detection}

Object detection models like Faster R-CNN~\cite{ren2015faster} that are two-stage networks including a region proposal network and a Fast R-CNN~\cite{girshick2015fast} work well but are slow.

One-shot object detectors like YOLO~\cite{redmon2016you}, SSD~\cite{liu2016ssd}, SqueezeDet~\cite{wu2017squeezedet} and DetectNet~\cite{tao2016detectnet} are faster and more suitable for mobile. They require only a single pass through the neural network using fixed grid detectors that allow the prediction of all bounding boxes at the same time. Those models are composed of a body network, usually pre-trained on a large image classification dataset like ImageNet, that acts as a feature extractor and a head network that detects objects.
% JPC: maybe citing these is too much for this? such as AlexNet~\cite{krizhevsky2012imagenet}, ResNet~\cite{he2016deep} and Inception~\cite{szegedy2015going}. 

However, those networks can be too large for mobile deployment. Instead, it will be preferred to use an architecture specifically designed for mobile like SqueezeNet~\cite{iandola2016squeezenet} that uses fire modules, SEP-NET~\cite{li2017sep} that uses filter convolutions and pattern residual blocks, MobileNet V1~\cite{howard2017mobilenets} that is based on depth-wise separable convolutions or MobileNetV2~\cite{sandler2018mobilenetv2} that adds linear bottleneck and expansion convolution to MobileNetV1\cite{Howard2017MobileNetsEC}.
%Then, on top of the feature extractor base network the object detection network has additional convolution layers that are fine-tuned to predict bounding boxes and class probabilities of the objects.
SDDLite~\cite{sandler2018mobilenetv2} has also been proposed as an object detection model and uses separable convolutions.

%Apps like EnvisionAI use a modified version of MobileNet for on-device object localization and detection. BlindTool and others design their own models that are trained on open image datasets like MS COCO~\cite{lin2014microsoft}, Google's Open Images~\cite{openimages2018}, or Pascal VOC~\cite{Everingham10}, and/or use a pre-trained network. 

\subsection{Model Compression}
Over the past few years deep neural networks have received attention for their state-of-the-art accuracy on tasks in computer vision and natural language processing, but these models are often too heavy and energy-hungry making them impossible to run or are very slow on mobile device. Therefore, strategies for model compression in mobile device have been developed into a large area of research.

For example, Deep Compression proposed by Han et al. (2015)~\cite{han2015deep} reduces the size of a network using 3 independent processes: pruning, trained quantization and Huffman encoding. 
Pruning transforms the original dense network into a sparse one by applying a certain threshold on the weight values to retain only the relevant connections. %This could be problematic as it can results in a lot of fine-tuning.
Once the network is compressed it can be fine-tuned. Low-rank factorization is a computationally intensive compression method that uses tensor decomposition to determine the relevant parameters~\cite{denton2014exploiting}.
Trained quantization consists in~\cite{han2015deep} of applying clustering over the original or pruned weights of the network. Then, the weights are replaced with the centroids of their cluster permitting weight sharing. 
%As for pruning, the compressed network is then fine-tuned.
Other methods of quantization exist including neural network binarization~\cite{courbariaux2016binarized} and trained ternary quantization~\cite{zhu2016trained}. The major drawback of these methods is that they are based on simple matrix approximations that can result in a reduction of accuracy.
Huffman encoding is a loss-less data compression technique using sparse matrix indices~\cite{han2015deep}.

Knowledge distillation can also be used to compress deep models. Those approaches shift knowledge from a larger teacher network into a smaller student network by learning the output class distributions using a softened softmax~\cite{bucilua2006model,romero2014fitnets}. 
Hinton et al. (2015)~\cite{hinton2015distilling} also proposed an ensemble of models composed of one or more full models and many specialist models which learn to distinguish fine-grained classes that the full models confuse.
However, those models can only be applied to classification through a softmax loss function.

%However, for all of those methods, a trade-off is to made between having a light model and keeping good performance.

\subsection{Federated Learning}
Federated learning is a way of training that allows for learning from data stored on a swarm of mobile devices without sacrificing privacy. In federated learning systems, a subset of users can identify a failing class, label training images, retrain their model locally, and share their improved weights to a central server where a federated learning algorithm produces a better, averaged model~\cite{mcmahan2016communication}. 

Models tailored for the visually impaired can be created through the use of federated learning systems. As previous research has explored~\cite{macleod2017understanding, salisbury2017toward}, the needs of blind users are fundamentally different than those of sighted ones as they do not have means of knowing the accuracy of a given description. 
We are aware of only one public visual question answering dataset, VizWiz~\cite{gurari2018vizwiz}, intended to assist the visually impaired community. More subtly problematic is the disparity between the image captioning datasets used to train these models, such as MS COCO~\cite{lin2014microsoft}, and the needs of this community who require descriptions that contain specific details. 
Federated learning allows user communities to retrain their models selecting exactly the kind of tasks they want to improve without sharing their private images to a server.

%In addition, we expect information that is relevant to one VIB user to be relatively important to another VIB user. Through the sharing of the model weights, users will implicitly have the ability to share their experiences resulting in models that are useful and relevant to many. 

%The solution would be realized if a subset of users identified the failing class, labelled a few images of the failing bill locally, retrained their model, and shared the improved weights to a central server. Interested users could download the new model at their convenience (and over wifi).
%While NantMobile's model is local, thereby not requiring an internet connection, a possible solution to removes the developer bottleneck would be to use federated learning~\cite{mcmahan2016communication}. 

\section{Where should the field go next?}
%Discuss proposed solutions for the VIB taking into account previously discussed technical details and usecases.
%The processing to accomplish tasks like text extraction, color identification, ambient light sensing, and object recognition have largely moved to the mobile device. The advantages of on-device processing are numerous, including increased performance from lower latency, increased privacy as the data never leaves the user's device, and independence from unreliable mobile coverage. 

Micro-navigation~\cite{Einsiedler2012IndoorMN} is an active area of research but has not yet matured. Existing solutions require models unsuitable for use on edge devices or specialized sensing hardware. Fundamental research on vision-and-language navigation~\cite{anderson2018vision}, outdoor navigation~\cite{Foell:2014:MUB:2694768.2694770}, scene segmentation~\cite{Alvarez:2012:RSS:2403272.2403301}, and visual question answering~\cite{Antol2015VQAVQ} are necessary before these models can deliver value to VIB people.

Advances in computer vision and natural language models, novel model-compression techniques, and increasing mobile processing power are already enabling the first on-device solutions for content summarization and navigation. Tasks like text extraction, color identification, ambient light sensing, and object recognition are being performed on-device to take advantage of increased performance from lower latency, better data privacy, and independence from unreliable mobile coverage; tasks that still require larger models and more processing power must offload the user’s data and processing to powerful servers which reduces autonomy and incurs service fees. As technologies improve and model accuracies increase, it is important to continue pushing processing towards edge devices to take advantage of their inherent gains in latency, privacy, and accessibility.

%Heavily compressed versions of hybrid convolutional and long-short term memory models for scene description \cite{Vinyals2017ShowChallenge} have begun to appear in apps like BlindSight. 

% \begin{itemize}
%     \item content summarization
%     \item winner-take-all unified app or federated services?
%     \item model compression for large networks (NLP)
%     \item A methodology for benchmarking performance of assistive technology.
% \end{itemize}

\small
% \bibliography{biblio}
\bibliographystyle{plain-nopagenum}

\end{document}